\documentclass[conference]{IEEEtran}
\IEEEoverridecommandlockouts
\usepackage{cite}
\usepackage{amsmath,amssymb,amsfonts}
\usepackage{algorithmic}
\usepackage{graphicx}
\usepackage{textcomp}
\usepackage{tabularx}
\usepackage{listings}
\usepackage{enumitem} 

\usepackage[hidelinks]{hyperref}
\usepackage{booktabs}
\usepackage{cleveref}
\usepackage{xcolor}
\usepackage{comment}
\usepackage{amssymb}
\usepackage[nolist,nohyperlinks]{acronym}

\usepackage{dirtytalk}
\usepackage{tikz}
\usepackage{pgfplots}
\usepackage{subcaption}
\usepackage{fontawesome5}
\usepackage{relsize}
\usepackage{colortbl}
\usepackage{caption}
\usepackage{subcaption}
\usepackage{listofitems} 
\usepackage{siunitx}

\usetikzlibrary{arrows, positioning, shapes.geometric, chains, arrows.meta, math}
\usetikzlibrary{circuits.logic.US}
\usetikzlibrary{circuits.logic.IEC}
\usetikzlibrary{decorations.pathreplacing, matrix, fit}
\usetikzlibrary{calc,shapes,shadows,patterns}
\usetikzlibrary{patterns, backgrounds}
\usetikzlibrary{pgfplots.groupplots}
\usepgfplotslibrary{groupplots, units}
\usetikzlibrary{3d,decorations.text,shapes.arrows,backgrounds}

\usepackage{circuitikz}

\usepackage{todonotes}
\usepackage{multirow}
\pgfplotsset{compat=1.18}
\definecolor{lightred}{RGB}{191,0,0}

\definecolor{UzLGray}{RGB}{194,217,230}
\definecolor{UzLBlue}{RGB}{0,75,90}
\definecolor{UzLBlue_medium}{RGB}{0,97,122}
\definecolor{UzLBlue_light_medium}{RGB}{0,145,168}
\definecolor{UzLBlue_light}{RGB}{0,174,199}

\definecolor{UzLOrange}{RGB}{236,116,4}
\definecolor{UzLRed}{RGB}{228,32,50}
\definecolor{UzLGreen}{RGB}{149,188,14}
\definecolor{UzLCyan}{RGB}{59,178,160}
\definecolor{UzLSky}{RGB}{110,165,206}

\let\oldlangle\langle
\let\oldrangle\rangle
\renewcommand{\langle}{\left\oldlangle}
\renewcommand{\rangle}{\right\oldrangle}

\tikzmath{
    \regDx = 3.2;
    \regDy = 1.5;
    \regTriangOff = \regDy / 8;
    \regTriangLen = (\regDy - \regTriangOff * 2) / 2;
    \angleDist = .1*\regDx; 
}

\tikzset{
    angular path right/.style args={#1}{ 
        to path={
            (\tikztostart) -| ($ (\tikztostart)!#1!(\tikztotarget) $) |- (\tikztotarget) \tikztonodes
        }
    },
    angular path left/.style args={#1}{ 
        to path={
            (\tikztostart) |- ($ (\tikztostart)!#1!(\tikztotarget) $) -| (\tikztotarget) \tikztonodes
        }
    },
    ScanChainS/.style={
        -stealth,
        thick,
        UzLGreen,
        densely dashed,
    }, 
    InBetweenConnect1S/.style={
        -stealth,
        thick,
        UzLCyan,
        densely dashed,
    },
    InBetweenConnect2S/.style={
        -stealth,
        thick,
        UzLRed,
        densely dashed,
    }, 
    InBetweenConnect3S/.style={
        -stealth,
        thick,
        UzLOrange,
        densely dashed,
    },
    InBetweenConnect4S/.style={
        -stealth,
        thick,
        UzLCyan,
        densely dashed,
    },
    InBetweenConnect5S/.style={
        -stealth,
        thick,
        lime,
        densely dashed,
    },
    DefaultConnectS/.style={
        black, 
        thick
    },
    InsideConnect1S/.style={
        UzLGreen, 
        thick
    },
    InsideConnect2S/.style={
        UzLRed,
        thick
    },
    InsideConnect3S/.style={
        UzLCyan,
        thick
    },
    InsideConnect4S/.style={
        UzLCyan,
        thick
    }
}

\tikzset{
    mux21/.style={
        muxdemux,
        muxdemux def={Lh=3.2, Rh=2.2, NL=2, NB=1, NR=1, w=1.8}, 
        muxdemux label={L1 = 0, L2 = 1}
    }, 
    RegS/.style={
        very thin,
        draw = black
    },
    pics/BoxP/.style args={#1} {
        code = {
            \draw [RegS] (0, 0) rectangle ++(\regDx,-\regDy); 
            \node [anchor=north west, align=center, minimum width = \regDx cm, minimum height = \regDy cm] {#1}; 
        }
    },
    pics/RegP/.style args={#1}{
        code = {
            \draw [RegS] (0, 0) rectangle ++(\regDx,-\regDy); 
            \draw [RegS] (0, -\regTriangOff) -- ++(\regTriangLen, -\regTriangLen) -- ++(-\regTriangLen, -\regTriangLen); 
            \node [anchor=north west, align=center, minimum width = \regDx cm, minimum height = \regDy cm] {#1}; 
        }
    },
    pics/RegEnP/.style args={#1}{
        code = {
            \pic {RegP={#1}};
            \node [RegS, rectangle, anchor=south] at (1.5*\regTriangLen, -\regDy) {en}; 
        }
    }
}

\usepackage{listofitems}
\tikzstyle{mynode}=[thick,draw=black,fill=UzLBlue,circle,minimum size=22]

\tikzset{pics/fake box/.style args={
#1 with dimensions #2 and #3 and #4}{
code={
\draw[ultra thin,fill=#1]  (0,0,0) coordinate(-front-bottom-left) to
++ (0,#3,0) coordinate(-front-top-right) --++
(#2,0,0) coordinate(-front-top-right) --++ (0,-#3,0) 
coordinate(-front-bottom-right) -- cycle;
\draw[ultra thin,fill=#1] (0,#3,0)  --++ 
 (0,0,#4) coordinate(-back-top-left) --++ (#2,0,0) 
 coordinate(-back-top-right) --++ (0,0,-#4)  -- cycle;
\draw[ultra thin,fill=#1!80!black] (#2,0,0) --++ (0,0,#4) coordinate(-back-bottom-right)
--++ (0,#3,0) --++ (0,0,-#4) -- cycle;
}
}}
\tikzset{pics/empty fake box/.style args={
#1 with dimensions #2 and #3 and #4}{
code={
\draw[ultra thin,fill=#1]  (0,0,0) coordinate(-front-bottom-left) to
++ (0,#3,0) coordinate(-front-top-right) --++
(#2,0,0) coordinate(-front-top-right) --++ (0,-#3,0) 
coordinate(-front-bottom-right) -- cycle;
\draw[ultra thin,fill=#1] (0,#3,0)  --++ 
 (0,0,#4) coordinate(-back-top-left) --++ (#2,0,0) 
 coordinate(-back-top-right) --++ (0,0,-#4)  -- cycle;
\draw[ultra thin,fill=#1!80!black] (#2,0,0) --++ (0,0,#4) coordinate(-back-bottom-right)
--++ (0,#3,0) --++ (0,0,-#4) -- cycle;
}
}}

\newif\ifcuboidshade
\newif\ifcuboidemphedge

\tikzset{
  cuboid/.is family,
  cuboid,
  shiftx/.initial=0,
  shifty/.initial=0,
  dimx/.initial=3,
  dimy/.initial=3,
  dimz/.initial=3,
  scale/.initial=1,
  densityx/.initial=1,
  densityy/.initial=1,
  densityz/.initial=1,
  rotation/.initial=0,
  anglex/.initial=0,
  angley/.initial=90,
  anglez/.initial=225,
  scalex/.initial=1,
  scaley/.initial=1,
  scalez/.initial=0.5,
  front/.style={draw=black,fill=white},
  top/.style={draw=black,fill=white},
  right/.style={draw=black,fill=white},
  shade/.is if=cuboidshade,
  shadecolordark/.initial=black,
  shadecolorlight/.initial=white,
  shadeopacity/.initial=0.15,
  shadesamples/.initial=16,
  emphedge/.is if=cuboidemphedge,
  emphstyle/.style={thick},
}

\usepackage{graphicx}

\def\BibTeX{{\rm B\kern-.05em{\sc i\kern-.025em b}\kern-.08em
    T\kern-.1667em\lower.7ex\hbox{E}\kern-.125emX}}

\begin{document}

\title{\textsc{Nail}: Not Another Fault-Injection Framework for Chisel-generated RTL
\thanks{This paper was supported in part by the ISOLDE project, nr. 101112274. ISOLDE is supported by the Chips Joint Undertaking and its members Austria, Czechia, France, Germany, Italy, Romania, Spain, Sweden, Switzerland. In addition, the paper was supported in part by projects: CeCaS (16ME0819) and RILKOSAN (16KISR010K) funded by Federal Ministry of Research, Technology and Space (BMFTR).}
}

\author{\IEEEauthorblockN{Robin Sehm, Christian Ewert, Rainer Buchty, Mladen Berekovic and Saleh Mulhem  }
\IEEEauthorblockA{\textit{Institute of Computer Engineering, }\textit{Universität zu Lübeck,} Lübeck, Germany\\
\{r.sehm, christian.ewert, rainer.buchty, mladen.berekovic, saleh.mulhem\}@uni-luebeck.de\
}
}

\maketitle
\begin{abstract}
Fault simulation and emulation are essential techniques for evaluating the dependability of integrated circuits, enabling early-stage vulnerability analysis and supporting the implementation of effective mitigation strategies.
High-level hardware description languages such as Chisel facilitate the rapid development of complex fault scenarios with minimal modification to the design. However, existing Chisel-based fault injection (FI) frameworks are limited by coarse-grained, instruction-level controllability, restricting the precision of fault modeling.

This work introduces \textsc{Nail}, a Chisel-based open-source FI framework that overcomes these limitations by introducing state-based faults. This approach enables fault scenarios that depend on specific system states, rather than solely on instruction-level triggers, thereby removing the need for precise timing of fault activation. For greater controllability, \textsc{Nail} allows users to arbitrarily modify internal trigger states via software at runtime.
To support this, \textsc{Nail} automatically generates a software interface, offering straightforward access to the instrumented design. This enables fine-tuning of fault parameters during active fault injection campaigns—a feature particularly beneficial for FPGA emulation, where synthesis is time-consuming.
Utilizing these features, \textsc{Nail} narrows the gap between the high speed of emulation-based FI frameworks, the usability of software-based approaches, and the controllability achieved in simulation.
We demonstrate \textsc{Nail}'s state-based fault injection and software framework by modeling a faulty general-purpose register in a RISC-V processor. Although this might appear straightforward, it requires state-dependent fault injection and was previously impossible without fundamental changes to the design.
The approach was validated in both simulation and FPGA emulation, where the addition of \textsc{Nail} introduced less than 1\% resource overhead.
\end{abstract}


\section{Introduction}\label{sec:Intro}
The trustworthiness of integrated circuits (ICs) is crucial, encompassing key design goals such as security, safety, and reliability~\cite{10530017}. Although there is no universally accepted definition of a trustworthy IC~\cite{bauer2022dependability, avizienis2004basic}, fault injection and fault propagation analysis are recognized as cross-domain concerns that significantly contribute to both dependability and security. Accordingly, there is a persistent demand for accurate and dynamic fault-evaluation tools, simulators, and emulators to support the development of trustworthy ICs~\cite{ayache2024holistic}.

Technology scaling has negatively impacted system dependability by increasing susceptibility to radiation-induced faults. This trend is particularly problematic for security- and safety-critical applications, such as aerospace and automotive systems. For instance, ISO\,26262 emphasizes the importance of fault injection approaches for validating automotive system-safety requirements~\cite{iso14a}.

Simulation environments that replicate and abstract integrated circuit architectures enable the deployment of fault models, supporting vulnerability analysis, reliability assessment, and verification. Fault simulation and emulation on FPGAs provide valuable insights into the effects of faults on digital circuits. The speed and accuracy of such simulations depend on the targeted level of abstraction~\cite{Benso2003FaultIT}.

Chisel~\cite{bachrach2012chisel} is a Scala-based high-level hardware description language (HDL) that introduces object-oriented and functional programming paradigms to digital circuit design. Chisel's extensible lowering process allows for transformation from Scala to intermediate representations like FIRRTL~\cite{firrtl1, firrtl2} and down to low-level HDLs such as Verilog. This extensibility enables fault simulation and emulation early in the design process with minimal modification from hardware designers. As a result, vulnerabilities can be identified and mitigated earlier and more efficiently.
This makes Chisel a suitable HDL for developing fault injection (FI) frameworks.

One notable representative of Chisel-based FI frameworks is \textit{Chiffre}~\cite{eldridge2018chiffre}. It leverages Chisel’s flexible backend to support injection of faults at the register-transfer level (RTL) with minimal design modification. To use it, one first annotates components of interest as faulty. \textit{Chiffre} then generates a hardware design with fault injection capabilities. 
To configure fault parameters such as masks, delays, or probabilities during runtime, \textit{Chiffre} introduced a software-controlled coprocessor that uses a standardized instruction-level interface. This interface is also used to trigger the fault. However, controllers based on instruction-level granularity—including the one provided by \textit{Chiffre}—are fundamentally limited: many fault scenarios cannot be accurately or feasibly modeled at this level of control.

To overcome these limitations, we present \textsc{Nail}, a next-generation open-source FI framework for Chisel-based circuits. As summarized in Table~\ref{tab:Comp}, \textsc{Nail} extends \textit{Chiffre} by supporting runtime-configurable, system-state-based fault triggers and provides a software framework for interfacing with the instrumented design. Importantly, this enhanced controllability allows modeling faults in components that were previously challenging to target, such as on-chip memories. These are frequently implemented as black boxes and are crucial for assessing system dependability~\cite{nicolaidis2010soft}.

\renewcommand{\arraystretch}{1.2}
\begin{table}[htbp]
\caption{Comparison of Fault Injection Frameworks}
\label{tab:Comp}
\footnotesize
\centering
\newcolumntype{L}{>{\raggedright\arraybackslash}X}
\begin{tabularx}{\columnwidth}{LLL}
\hline
\textbf{Feature} & \textbf{\textit{Chiffre}~\cite{eldridge2018chiffre}} & \textbf{\textsc{Nail}} \\ \hline
\textbf{Trigger flexibility} & 
\begin{itemize}[leftmargin=*, label={-}, itemsep=0pt, parsep=0pt, topsep=0pt, partopsep=0pt]
    \item Instruction-level
\end{itemize} &
\begin{itemize}[leftmargin=*, label={-}, itemsep=0pt, parsep=0pt, topsep=0pt, partopsep=0pt]
    \item State-dependent triggers
    \item Runtime-configurability of triggers 
    \item Instruction-level
\end{itemize} \\ \hline
\textbf{Configuring fault injection campaigns} &
\begin{itemize}[leftmargin=*, label={-}, itemsep=0pt, parsep=0pt, topsep=0pt, partopsep=0pt]
    \item Manual configuration through command-line and JSON files
    \item Limited runtime reconfiguration
\end{itemize} &
\begin{itemize}[leftmargin=*, label={-}, itemsep=0pt, parsep=0pt, topsep=0pt, partopsep=0pt]
    \item Flexible configuration through command-line, C++ interface, or JSON files
    \item Supports both compile-time and runtime reconfiguration
\end{itemize} \\ \hline
\textbf{Granularity of injector control} &
\begin{itemize}[leftmargin=*, label={-}, itemsep=0pt, parsep=0pt, topsep=0pt, partopsep=0pt]
    \item Only support enabling or disabling all injection points at once
\end{itemize} &
\begin{itemize}[leftmargin=*, label={-}, itemsep=0pt, parsep=0pt, topsep=0pt, partopsep=0pt]
    \item Fine-grained control: enable or disable injection points independently
\end{itemize} \\ \hline
\textbf{Targeting of multi-signal structures (e.g., bundles)} &
\begin{itemize}[leftmargin=*, label={-}, itemsep=0pt, parsep=0pt, topsep=0pt, partopsep=0pt]
    \item Fault injection applies to entire structure only
\end{itemize} &
\begin{itemize}[leftmargin=*, label={-}, itemsep=0pt, parsep=0pt, topsep=0pt, partopsep=0pt]
    \item Can target any subset of signals within a structure
\end{itemize} \\ \hline
\end{tabularx}
\end{table}

\subsection{Paper Contribution}
The main contributions of this work are as follows:
\begin{itemize}
    \item Introduction of system-state-based and runtime-configurable triggers for fault injection.
    \item Implementation of the \textsc{NailCompanion}, an auto-generated software interface that enables user-friendly fault configuration at compile time and runtime.
    \item Increased flexibility in specifying fault injection targets.
    \item Enables targeting any subset of signals bundles, increasing precision and scalability instead of limiting fault injection targets to only complete bundles.
\end{itemize}
Despite these improvements, \textsc{Nail} introduces minimal hardware overhead and requires no extensive modifications to the Chisel design.

\subsection{Structure}
The rest of this paper is structured as follows: Section~\ref{sec:related_work} provides an overview of related FI frameworks and discusses our motivation. Section~\ref{sec:methodology} describes the proposed \textsc{Nail} framework and its key components. Section~\ref{sec:exper} validates our approach by describing a fault injection performed on a RISC-V core. Finally, Section~\ref{sec:con} summarizes our findings and concludes the paper.
\section{Background \& Motivation}
\label{sec:related_work}

Fault injection frameworks are essential for evaluating the robustness and dependability of hardware designs. These frameworks can be classified into four main approaches, each with distinct trade-offs in observability, controllability, and required engineering effort:

\begin{itemize}
    \item [(1)] \textbf{Hardware-based Fault Injection} applies faults to fabricated ASICs through dedicated hardware, allowing non-intrusive FI experiments with real silicon. 
    Provided, no scan chain is included in the design, this approach is limited to pin-level observability and controllability, restricting the precision of fault analysis~\cite{karlsson1998application, madeira1994rifle}.
    
    \item[(2)] \textbf{Software-based Fault Injection} introduces faults during software execution, either at compile time~\cite{Thomas2013LLFIA} or runtime~\cite{Carreira2001Xception}. This method improves flexibility and ease of use, requiring little hardware knowledge, but is typically confined to the processor's register or instruction level, offering limited insight into the effects of faults on internal hardware structures.
    \item[(3)] \textbf{Simulation-based Fault Injection} provides fine-grained control and detailed visibility by introducing faults at the register, flip-flop, or gate level. This enables comprehensive analysis across various abstraction levels~\cite{Bekele2023SurveyQEMU, sieh1997verify, mansour2013automated}, but comes with high computational overhead and demands detailed hardware design knowledge.
    \item[(4)] \textbf{Emulation-based Fault Injection} accelerates fault analysis by deploying the designs onto FPGAs. This approach can achieve near real-time speed and fine-grained fault injection by instrumenting the hardware with scan chains and dedicated controllers~\cite{Nowosielski2015FLINT}. However, existing emulation-based approaches are typically limited in trigger flexibility, often relying on simple, externally-controlled events or instruction-level triggers. 
\end{itemize}

Frameworks such as \textit{Chiffre}~\cite{eldridge2018chiffre} have advanced emulation-based fault injection by automating much of the instrumentation process. \textit{Chiffre} enables fault injection to be configured at runtime with minimal manual changes, lowering the barrier for early-stage dependability analysis. Nevertheless, it remains limited by its reliance on instruction-level triggers and lacks support for complex, state-dependent fault scenarios and a comprehensive software interface.

\textsc{Nail} addresses these limitations by combining the strengths of the aforementioned fault injection methodologies into a unified framework. It provides the ease of use characteristic of software-based approaches (2) through its auto-generated C++ framework. At the same time, \textsc{Nail} achieves the fine-grained control and arbitrarily complex trigger capabilities found in advanced simulation-based methods (3), as well as the high-speed analysis made possible by emulation-based approaches (4) via its native integration with the Chisel.

As a result, \textsc{Nail} effectively balances controllability, observability, and development effort, narrowing the gap between these methodologies.
\section{Methodology} \label{sec:methodology}

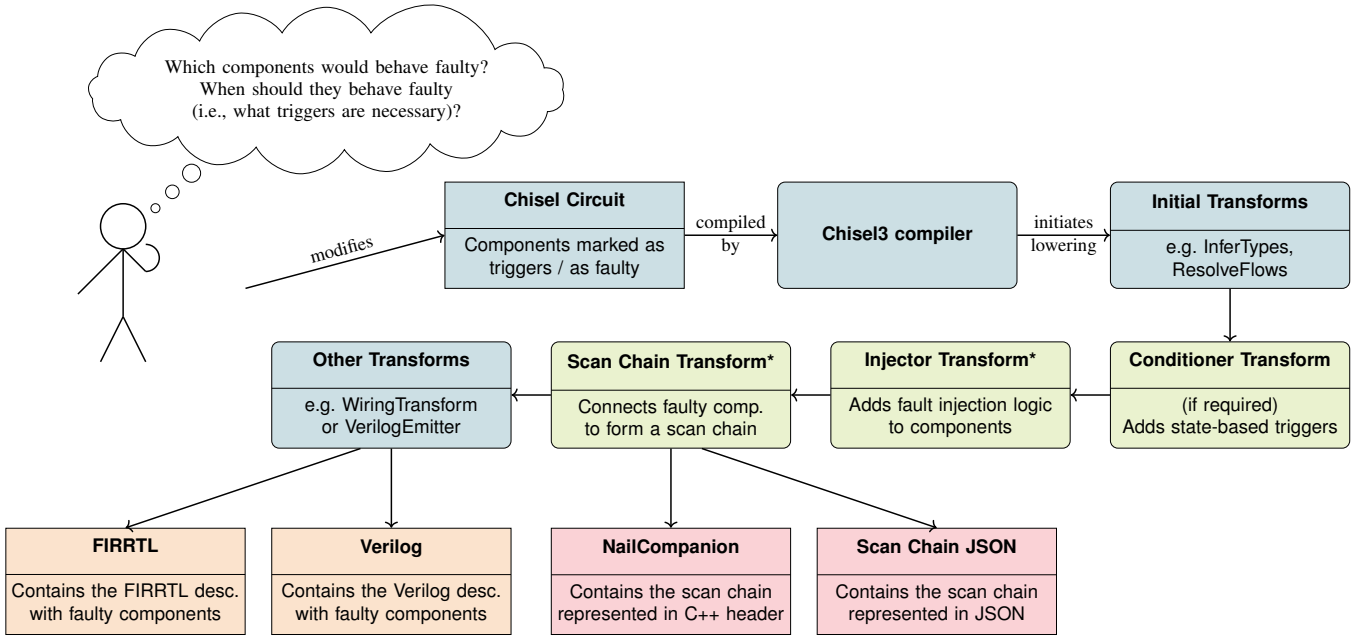
\begin{figure*}[htbp]
    \centering
    \resizebox{\textwidth}{!}{ %
        \begin{tikzpicture}[
    pics/UserP/.style={
        code = {
            \node[draw, thick, circle, minimum size=1cm] (head) at (0,0) {};
            \draw[thick] (0,-0.5) -- (0,-2);
            \draw[thick] (0,-0.5) -- (-1,-1.5); 
            \draw[thick] (0,-0.5) -- (0.5,-1); 
            \draw[thick] (0.5,-1) arc[start angle=-90, end angle=90, radius=0.3];
            \draw[thick] (0,-2) -- (-0.5,-3); 
            \draw[thick] (0,-2) -- (0.5,-3); 
            \node[draw, cloud, cloud puffs=16.6, cloud puff arc=120, aspect=4, fill=white, align=center, minimum width=1.8cm, minimum height=1cm, font=\large] (cloud) at (4.5,3) {Which components would behave faulty?\\When should they behave faulty\\(i.e., what triggers are necessary)?};
            \draw[fill=white] ($(head.20)!.18!(cloud.210)$) circle(0.1cm); 
            \draw[fill=white] ($(head.20)!.47!(cloud.210)$) circle(0.15cm); 
            \draw[fill=white] ($(head.20)!.8!(cloud.210)$) circle(0.2cm); 
            }
        },
    RegDescS/.style={
        draw, 
        rectangle, 
        minimum width=4cm, 
        minimum height=2cm, 
        align=center, 
        inner sep=0, 
        font=\sffamily 
        }
    ]
    \pic [local bounding box=user, transform shape, scale=.85, minimum height=4cm] at (0, 0) {UserP};
        
    \node[RegDescS, text width=4.5cm, fill=UzLBlue_medium!20] (chisel3Code) at ($(user.east) + (0, -1)$) {
        \textbf{Chisel Circuit}\\
        \hrulefill\\
        Components marked as\\triggers / as faulty
    };
    
    \coordinate (userHand) at ($(user.east) - (6, 2)$);

    \draw[DefaultConnectS, ->] (userHand) -- node[align=center, sloped, above] {modifies} (chisel3Code.west);

    \node[RegDescS, rounded corners, text width=4.5cm, fill=UzLBlue_medium!20] (chisel3Compiler) at ($(chisel3Code.east) + (4, 0)$) {
        \textbf{Chisel3 compiler}
    };

    \draw[DefaultConnectS, ->] (chisel3Code.east) -- node[align=center] {compiled\\by} (chisel3Compiler.west);

    \node[RegDescS, text width=4.5cm, rounded corners, fill=UzLBlue_medium!20] (initTf) at ($(chisel3Compiler.east) + (4, 0)$) {
        \textbf{Initial Transforms}\\
        \rule{\linewidth}{0.4pt}\\
        e.g. InferTypes,\\ResolveFlows
    };

    \draw[DefaultConnectS, ->] (chisel3Compiler.east) -- node[align=center] {initiates\\lowering} (initTf.west);

    \node[RegDescS, text width=4.5cm, rounded corners, fill=UzLGreen!20] (ConditionerTF) at ($(initTf.south) + (0, -2)$) {
        \textbf{Conditioner Transform}\\
        \rule{\linewidth}{0.4pt}\\
        (if required)\\Adds state-based triggers
    };

    \draw[DefaultConnectS, ->] (initTf.south) -- (ConditionerTF.north);

    \node[RegDescS, text width=4.5cm, rounded corners, fill=UzLGreen!20] (InjectorTF) at ($(ConditionerTF.west) + (-3, 0)$) {
        \textbf{Injector Transform*}\\
        \rule{\linewidth}{0.4pt}\\
        Adds fault injection logic\\ to components
    };

    \draw[DefaultConnectS, ->] (ConditionerTF.west) -- (InjectorTF.east);

    \node[RegDescS, text width=4.5cm, rounded corners, fill=UzLGreen!20] (ScanChainTF) at ($(InjectorTF.west) + (-3, 0)$) {
        \textbf{Scan Chain Transform*}\\
        \rule{\linewidth}{0.4pt}\\
        Connects faulty comp.\\
        to form a scan chain
    };

    \draw[DefaultConnectS, ->] (InjectorTF.west) -- (ScanChainTF.east);

    \node[RegDescS, text width=4.5cm, rounded corners, fill=UzLBlue_medium!20] (OtherTF) at ($(ScanChainTF.west) + (-3, 0)$) {
        \textbf{Other Transforms}\\
        \rule{\linewidth}{0.4pt}\\
        e.g. WiringTransform\\
        or VerilogEmitter
    };

    \draw[DefaultConnectS, ->] (ScanChainTF.west) -- (OtherTF.east);

    \node[RegDescS, text width=4.5cm, fill=UzLRed!20] (NailCompanion) at ($(ScanChainTF.south) + (0, -2.5)$) {
        \textbf{NailCompanion}\\
        \rule{\linewidth}{0.4pt}\\
        Contains the scan chain\\
        represented in C++ header
    };

    \draw[DefaultConnectS, ->] (ScanChainTF.270) -- (NailCompanion.north);

    \node[RegDescS, text width=4.5cm, fill=UzLRed!20] (scanChainJson) at ($(ScanChainTF.south) + (5, -2.5)$) {
        \textbf{Scan Chain JSON}\\
        \rule{\linewidth}{0.4pt}\\
        Contains the scan chain\\
        represented in JSON
    };

    \draw[DefaultConnectS, ->] (ScanChainTF.300) -- (scanChainJson.north);

    \node[RegDescS, text width=4.5cm, fill=UzLOrange!20] (firrtlFaulty) at ($(OtherTF.south) + (-5, -2.5)$) {
        \textbf{FIRRTL}\\
        \rule{\linewidth}{0.4pt}\\
        Contains the FIRRTL desc.\\
        with faulty components
    };

    \draw[DefaultConnectS, ->] (OtherTF.240) -- (firrtlFaulty.north);

    \node[RegDescS, text width=4.5cm, fill=UzLOrange!20] (verilog) at ($(OtherTF.south) + (0, -2.5)$) {
        \textbf{Verilog}\\
        \rule{\linewidth}{0.4pt}\\
        Contains the Verilog desc.\\
        with faulty components
    };

    \draw[DefaultConnectS, ->] (OtherTF.270) -- (verilog.north);

\end{tikzpicture}
    }
    \caption{
        Tool flow for generating a faulty design with \textsc{Nail}. The process begins by modifying the Chisel circuit. Stages are color-coded for clarity: blue indicates standard Chisel stages, green denotes stages from \textsc{Nail}, orange highlights the lowered Verilog and FIRRTL hardware descriptions (including faulty components), and red represents the serialized scan chain configuration. Transforms marked (*) are adapted from \textit{Chiffre}.
}\label{fig:NailToolflow}
\end{figure*}

This section presents \textsc{Nail}\footnote{\url{https://gitlab.iti.uni-luebeck.de/pubs/nail}}, a next-generation fault injection framework designed for Chisel-based circuit designs. Fig.~\ref{fig:NailToolflow} illustrates the tool flow, demonstrating how \textsc{Nail} facilitates the generation of both hardware descriptions with integrated fault injection capabilities and the corresponding software required to conduct fault injection experiments on the instrumented hardware.

To enhance fault observability, the generated Verilog can be simulated using tools such as Verilator\cite{githubGitHubVerilatorverilator}. During simulation, \textsc{Nail} produces detailed logs of each injected fault which aids in refining trigger conditions and adjusting fault parameters.
Alternatively, when execution speed is prioritized over observability, the Verilog description can be synthesized and deployed on an FPGA for fault emulation.

The software controlling the fault injection experiment (the \textit{driver code}) must be tailored to the specific circuit under test. Section~\ref{sec:exper} provides an example using \textsc{Nail}'s software framework, where the generated C++ header is included in the driver code, compiled, and deployed on a RISC-V based SoC.

\subsection{Preliminaries}
As \textsc{Nail} uses many ideas presented in \textit{Chiffre}~\cite{eldridge2018chiffre}, we summarize its core concepts and the RISC-V-based coprocessor, \textit{leChiffre}. \textit{Chiffre} enables fault simulation and emulation for Chisel-based circuits. Users first annotate target components for fault injection in the Chisel design. Automated transformations then integrate the necessary hardware extensions and injectors directly into the circuit. The framework supports multiple injector types, enabling the modeling of both transient and permanent faults, as well as time-based fault scenarios. Note that the fault type is determined by the injector's hardware implementation and cannot be changed at runtime.\footnote{A combined injector with a runtime-selectable bit could allow dynamic fault type selection.}

An important architectural feature of \textit{Chiffre} (and many other FI frameworks) is the use of \textit{scan chains} and \textit{scan fields} to provide flexible and precise control over fault parameters at runtime. A scan chain is a sequence of interconnected registers called scan fields inserted into the hardware design. Each scan field typically corresponds to a configurable aspect of an injector, such as bit masks, flip probabilities, or target values. By organizing these scan fields into a chain, the system can efficiently configure and update these parameter for multiple components via a serial (the approach used in \textit{Chiffre}) or parallel interface. 
We refer to an assignment values for all scan fields in a scan chain as \textit{scan chain configuration}.

To manage fault injection campaigns, \textit{Chiffre} connects the scan chain of injectors to a dedicated controller, such as the RISC-V-based coprocessor \textit{leChiffre}. This controller can receive a new scan chain configuration via software through a standardized interface called RoCC\footnote{Rocket Custom Coprocessor}.

\begin{figure*}[htbp]
    \centering
    \resizebox{\textwidth}{!}{
        \tikzmath{
    \conditionerPadX = 1;
    \conditionerPadY = .5;
    \spaceDx = \regDx/5;
    \spaceDy = \regDy/5;
    \regDxShift = .5*\regDx;
    \regDyShift = .4*\regDy;
    \wireDx = \regDx - \regDxShift + 2*\angleDist;
    \wireDy = \regDy/2 + \conditionerPadY/2;
    \componentDx = 9.5;
    \componentDy = 10.5;
    \portInDy = \componentDy / 5;
    \portOutDy = \componentDy / 5;
    \outsideDx = 2.5;
}

\begin{tikzpicture}[
    pics/ConditionerP/.style={
        code = {
            \coordinate (condIoScanEn) at (0, -\portInDy);
            \coordinate (condIoCondIn) at (0, -2*\portInDy);
            \coordinate (condIoScanIn) at (0, -3*\portInDy);
            \coordinate (condIoCondEn) at (\componentDx, -1*\portOutDy);
            \coordinate (condIoCond1) at (\componentDx, -2*\portOutDy);
            \coordinate (condIoCond2) at (\componentDx, -3*\portOutDy);
            \coordinate (condIoScanOut) at (\componentDx, -4*\portOutDy);
            \coordinate (condCoordIsActive) at (\conditionerPadX, -4);
            \draw [RegS, rounded corners = .5cm] (0, 0) rectangle ++(\componentDx, -\componentDy);
            \pic [local bounding box=condRegEnabled] at (\conditionerPadX, -\conditionerPadY) {RegP=enabled};
            \pic [local bounding box=condRegIsActive] at (condCoordIsActive) {RegEnP=isActive};
            \pic [local bounding box=condRegSf1] at ($(condRegIsActive.south west) + (\regDxShift, -\regDyShift)$) {RegEnP=ScanField$_i$};
            \pic [local bounding box=condRegSf2] at ($(condRegSf1.south west) + (\regDxShift, -\regDyShift)$) {RegEnP=ScanField$_{i+1}$};
            \node[circuit logic US, and gate, draw, anchor=west, inputs={nn}] at ($(condIoCondEn) - (2, .5)$) (condGateAnd) {};
            \node[circuit logic US, or gate, draw, anchor=west, inputs={ni}] at ($(condIoCondEn) - (5, 1)$) (condGateOr) {};
            \draw[InsideConnect1S] (condRegIsActive.east) -- ++(\angleDist, 0) to [angular path left={.5}] ($(condRegSf1.west) - (\angleDist, 0)$) -- (condRegSf1.west);
            \draw[InsideConnect1S] (condRegSf1.east) -- ++(\angleDist, 0) to [angular path left={.5}] ($(condRegSf2.west) - (\angleDist, 0)$) -- (condRegSf2.west);
            \draw[InsideConnect1S] (condIoScanIn) to[angular path right={.5}] (condRegIsActive.west);
            \draw[InsideConnect1S] (condIoScanEn) to[angular path right={.5}] (condRegEnabled.west);
            \draw[InsideConnect4S] (condIoCondIn) to[angular path right={.2}] (condGateOr.input 1);
            \draw[InsideConnect4S] (condRegEnabled.east) to[angular path right={.3}] (condGateAnd.input 1);
            \draw[InsideConnect4S] (condRegIsActive.40) |- (condGateOr.input 2);
            \draw[InsideConnect1S] (condRegSf2.east) to[angular path right={.5}] (condIoScanOut);
            \draw[InsideConnect4S] (condRegSf1.30) |- (condIoCond1);
            \draw[InsideConnect4S] (condRegSf2.30) |- (condIoCond2);
            \draw[InsideConnect4S] (condGateOr.output) to[angular path right={.5}] (condGateAnd.input 2);
            \draw[InsideConnect4S] (condGateAnd.output) to[angular path right={.5}] (condIoCondEn);
        }
    }, 
    pics/CondInjectorP/.style={
        code = {
            \coordinate (injIoScanEn) at (0, -\portInDy);
            \coordinate (injIoScanIn) at (0, -3*\portInDy);
            \coordinate (injIoScanOut) at (\componentDx, -4*\portOutDy);
            \coordinate (injIoOut) at (\componentDx, -\portOutDy);
            \coordinate (injIoIn) at (.5*\componentDx, 0);
            \coordinate (injCoordIsActive) at (\conditionerPadX, -4);
            \draw [RegS, rounded corners = .5cm] (0, 0) rectangle ++(\componentDx, -\componentDy);
            \pic [local bounding box=injRegIsActive] at (injCoordIsActive) {RegEnP=isActive};
            \pic [local bounding box=injRegSf1] at ($(injRegIsActive.south west) + (\regDxShift, -\regDyShift)$) {RegEnP=ScanField$_i$};
            \pic [local bounding box=injRegSf2] at ($(injRegSf1.south west) + (\regDxShift, -\regDyShift)$) {RegEnP=ScanField$_{i+1}$};
            \node[circuit logic US, and gate, draw, anchor=west, inputs={nn}] at ($(injRegIsActive.north) + (0, .7)$) (injGateAnd) {};
            \node[cloud, cloud puffs=8.6, cloud ignores aspect, minimum width=3cm, minimum height=2cm, align=center, draw] (injMiscImpl) at ($(injRegIsActive.east) + (3.5, 0)$) {Injector\\Impl.};
            \node[mux21, align=center] (injMux) at ($(injIoOut) - (4, 0)$) {use\\faulty};
            \draw[InsideConnect1S] (injRegIsActive.east) -- ++(\angleDist, 0) to [angular path left={.5}] ($(injRegSf1.west) - (\angleDist, 0)$) -- (injRegSf1.west);
            \draw[InsideConnect1S] (injRegSf1.east) -- ++(\angleDist, 0) to [angular path left={.5}] ($(injRegSf2.west) - (\angleDist, 0)$) -- (injRegSf2.west);
            \draw[InsideConnect2S] (injRegIsActive.120) |- (injGateAnd.input 2);
            \draw[InsideConnect1S] (injIoScanIn) to[angular path right={.5}] (injRegIsActive.west);
            \draw[InsideConnect1S] (injRegSf2.east) to[angular path right={.5}] (injIoScanOut);
            \draw[InsideConnect2S, -stealth] (injRegSf1.30) |- (injMiscImpl.west);
            \draw[InsideConnect2S, -stealth] (injRegSf2.30) to[angular path left={.5}] (injMiscImpl.south);
            \draw[InsideConnect4S] (injIoScanEn.east) to[angular path right={.5}] (injGateAnd.input 1);
            \draw[InsideConnect2S] (injGateAnd.output) -| (injMux.bpin 1);
            \draw[InsideConnect2S] (injMiscImpl.north west) -- ++(-\angleDist, 0) -| ($(injMux.lpin 2) - (\angleDist, 0)$) -- (injMux.lpin 2);
            \draw[InsideConnect2S] (injIoIn) |- (injMux.lpin 1);
            \draw[InsideConnect2S] (injMux.rpin 1) to[angular path right={.2}] (injIoOut);
            \draw[InsideConnect2S] (injIoIn) to[angular path left={.2}] (injMiscImpl.north);
        }
    }
]
\pic [local bounding box=cond] at (4, -3) {ConditionerP};
\pic [local bounding box=inj] at ($(cond.north east) + (3, 0)$) {CondInjectorP};
\pic [local bounding box=fc] at ($(inj.north) + (-1.6, 2.8)$) {RegP=Faulty\\Component};

\node[cloud, cloud puffs=8.6, cloud ignores aspect, minimum width=2cm, 
minimum height=1.5cm, align=center, draw] 
(fcLogicLeft) at ($(fc) - (\regDx + 1.5, 0)$) {};

\node[cloud, cloud puffs=8.6, cloud ignores aspect, minimum width=2cm, 
minimum height=1.5cm, align=center, draw] 
(fcLogicRight) at ($(fc) + (\regDx + 2, 0)$) {};

\node[cloud, cloud puffs=20.6, cloud ignores aspect, minimum width=3cm, minimum height=2.2cm, align=center, draw] (conditionImpl) at ($(condRegEnabled) + (2, 3)$) {Condition:\\$f(\text{ScanField}_i, \text{ScanField}_{i+1}, \dots) \to \{0,1\}$};

\draw[InBetweenConnect2S] (fc.south) -- ++(0, -\angleDist) -| (injIoIn);
\draw[InBetweenConnect2S] (injIoOut) -- ($(injIoOut) + (\angleDist, 0)$) -| (fcLogicRight);
\draw[DefaultConnectS] (fc.east) -- (fcLogicRight) node[midway, align=center, font=\bfseries, fill=white] {\faTrash};
\draw[DefaultConnectS] (fcLogicLeft) -- (fc.west);

\draw[InBetweenConnect1S] (condIoCondEn) -- (injIoScanEn);
\draw[InBetweenConnect1S] (condIoCond1) -- ++(2*\angleDist, 0) |- (conditionImpl.-6);
\draw[InBetweenConnect1S] (condIoCond2) -- ++(4*\angleDist, 0) |- (conditionImpl.4);
\draw[InBetweenConnect1S] (conditionImpl.west) -| ($(condIoCondIn) - (2*\angleDist, 0)$) -- (condIoCondIn);

\draw[ScanChainS] (condIoScanOut) to[angular path right={.6}] (injIoScanIn);
\draw[ScanChainS] ($(condIoScanIn) - (\outsideDx, 0)$) -- (condIoScanIn);
\draw[ScanChainS] ($(condIoScanEn) - (\outsideDx, 0)$) -- (condIoScanEn);
\draw[ScanChainS] (injIoScanOut) -- ($(injIoScanOut) + (\outsideDx, 0)$);

\node[font=\Large, anchor=south west] at ($(cond.south west) + (.5, 1)$) {Conditioner};
\node[font=\Large, anchor=south west] at ($(inj.south west) + (.5, 1)$) {Injector};

\end{tikzpicture}
    }
    \caption{
    Overview of a conditioner and injector within a scan chain, as implemented in \textsc{Nail}. The blue circuit controls the condition logic, evaluating the output of a fixed Boolean function that takes the scan fields as input. The red circuit generates the faulty signal according to the injector's implementation, while the green circuit manages the scan chain control.
    }\label{fig:condInjector}
\end{figure*}
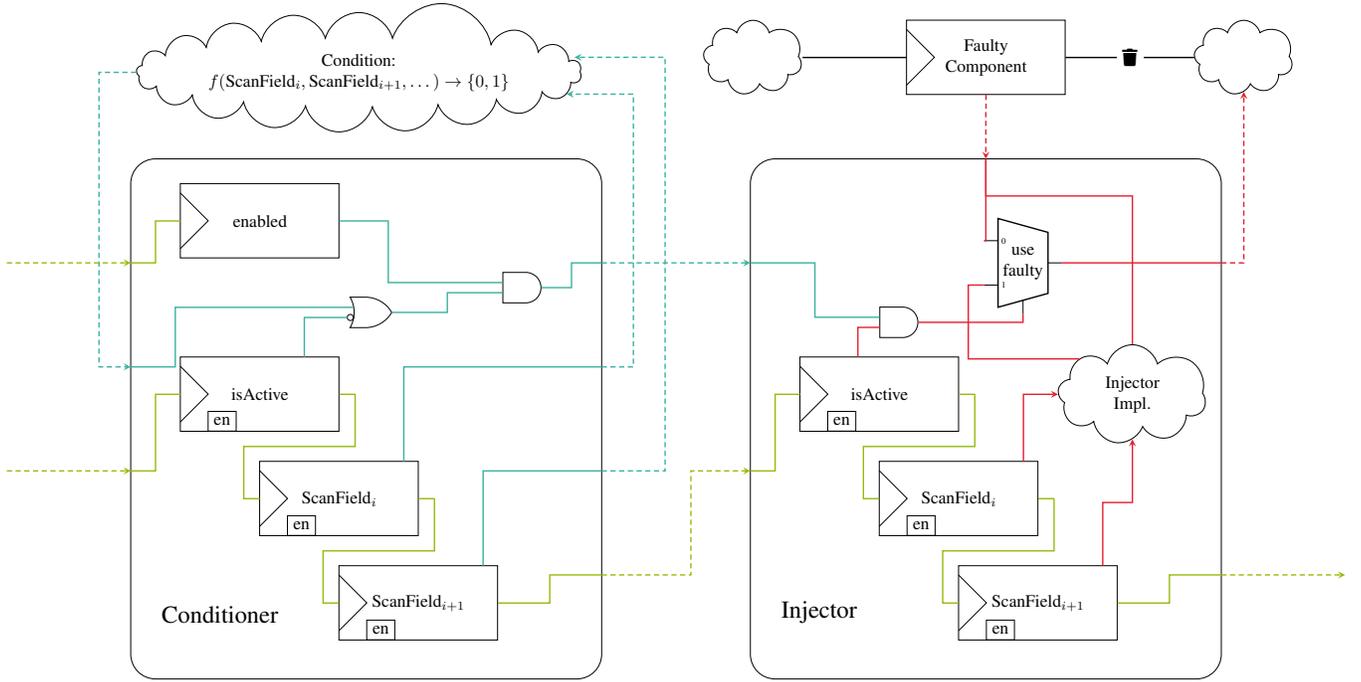

\subsection{Conditional Fault Injection}\label{sec:ConditionalFaultjection}
A major advancement of \textsc{Nail} is its support for conditional, system-state-based fault injection.
Unlike traditional approaches where faults are injected only at fixed times (\textit{instruction-level}), \textsc{Nail} allows faults to be triggered when user-defined conditions within the circuit are satisfied. That is, one or multiple components (i.e. signals) are evaluated to a single bit, indicating whether a fault should occur. We refer to this enable bit as the \textit{condition}. In \textsc{Nail}, the condition may depend on any number of scan fields placed anywhere in the circuit.

At a high level, these \textit{conditions} are modeled as Boolean functions parametrized by scan fields. The result of this function determines whether the associated fault injector is enabled.
\textsc{Nail} introduces the notion of a \textit{conditioner} to encapsulate this logic: a conditioner is a grouping of scan fields and associated logic that collectively define the condition under which a fault is injected, analogous to how an injector is a collection of scan fields and fault logic.
For example, let $SF_1, SF_2$ denote two scan fields in the conditioner and $X_1, X_2$ two components in the circuit. Then a condition $f$ can be obtained as follows:
\begin{equation*}
  f(SF_1, SF_2) = X_1 == SF_1 \lor X_2 == SF_2 \\
\end{equation*}
Here, $SF_1, SF_2$ can be set during runtime, and $X_1, X_2$ contain the \textit{state} of the circuit that is used as a trigger to inject a fault.

From the user's perspective, making a fault depend on an internal state is fairly straightforward. First, the condition including its scan fields is described within the targeted Chisel module. To pass it to \textsc{Nail}'s backend, a function call (\texttt{markAsFaulty}) is used.
It requires the faulty component and (optionally) the newly introduced condition and other parameters. See Fig.~\ref{fig:faultyRegfileScala} how it is implemented in Scala. The framework then ensures that the injector is only active when the condition evaluates to true. Note that in \textsc{Nail}, all components' \textit{isActive} state must be configured via the scan chain. Furthermore, the scan chain must be enabled by the controller for the fault to take effect. Fig.~\ref{fig:condInjector} illustrates the architecture, where a conditioner and its injector are placed within the scan chain.

\subsection{Integrating Condition Logic}
To support state-based fault injection, \textsc{Nail} automatically instruments the target circuit during the lowering process. This involves identifying annotated targets and conditions, and inserting the required logic to evaluate conditions and control injector activation.
The Chisel3-based transformation workflow operates on the circuit's intermediate representation (IR), ensuring that the necessary connections between condition logic, scan fields, and injectors are established without user intervention. The transformation identifies all scan fields in the IR that contribute to each condition. It then constructs and inserts the appropriate conditioner containing these scan fields into the circuit.

The \texttt{firrtl.passes.wiring.WiringTransform} is used to connect the scan fields in the conditioner (\textit{sources}) to their respective places (\textit{sinks}) inside the condition~\footnote{%
For this, Source- and SinkAnnotations are used. These are used to indicate where new connections should be created during lowering. A SourceAnnotation marks a signal (or port) as the origin for wiring, while a SinkAnnotation marks a destination. The WiringTransform uses these annotations to insert the necessary wires between sources and sinks automatically, simplifying circuit instrumentation.}.

By abstracting these mechanisms into automated transformations, \textsc{Nail} makes it possible to implement advanced, state-based fault injection scenarios without manual modification of the original hardware design.

\subsection{Software Framework}\label{sec:Companion}
Both \textit{Chiffre} and \textsc{Nail} offer a combined approach to generate new scan chain configurations. Both emit a JSON file during lowering that describes all scan fields present in the design, including their order and respective widths. Using a Scala-based command-line tool, the user can now define the values of the scan fields and emit a binary representation of the desired configuration.

For RoCC cores like \textit{leChiffre} the user may want to convert this binary representation into a C-style array for use within the driver code. However, relying on a JSON parser or Scala may be impractical when targeting low-power devices, when the design is simulated or deployed on bare-metal systems. Furthermore, rapid re-configuration, i.e. creating a new scan chain configuration during the fault injection experiment requires a very efficient implementation of the scan chain.

To address these issues, \textsc{Nail} introduces the \textsc{NailCompanion}. During lowering, it automatically generates a C++ header that describes the scan chains present in the hardware design.
In addition, it implements a readable and type-safe API to generate new scan chain configurations and to interact with the RoCC controller.
The header is generated by analyzing the scan chain layout during the lowering process. A combination of C++ \texttt{enum classes} and other modern C++ features ensure a readable link between the component names in the Chisel design and their representation in C++. 
The representation of a scan chain configuration is implement as a plain C-style array. All operations, i.e. setting the values of scan fields are done using bit manipulation.
Due to this efficient implementation, the resulting interface is very performant and abstracts away error-prone low-level array modification allowing users to configure scan fields using intuitive identifiers.

To use this header, it needs to be included in the driver code and compiled with a C++20-capable compiler. Fig.~\ref{fig:FaultyRegfileCpp} provides an example how to combine the driver code and scan chain configuration. To set a scan field using the \textsc{NailCompanion}, one must know its type and the name of the faulty component. For example, to set the value of a component that should behave \textit{permanently} faulty (\textit{stuck-at}), the type \texttt{nail::inject::StuckAt} can be used. When deployed on a RISC-V core, the header also provides a way to send RoCC commands directly to the scan chain controller.

\subsection{Modification to existing Work}
To enable \textsc{Nail}'s enhancements, all of \textit{Chiffre}'s transforms were rewritten and several components, such as the injector, were modified.
For example, the enable signal from the controller needs to be buffered in a register in order to reduce critical path length. However, since a (optional) conditioner already provides a buffer, the injector implementation needs to be aware whether a conditioner is attached or not so there is exactly one register buffering the enable signal.
As mentioned previously, all \textsc{Nail} components can be turned on or off via the scan field (\textit{isActive}). This facilitates the creation of designs that support multiple fault scenarios - injectors not needed for the fault injection experiment can just be deactivated. 
The final contribution, as summarized in Table~\ref{tab:Comp}, stems from improving fault injection into \texttt{chisel3.Bundle}s. These bundles provide a grouping of several signals to form a single entity. Many Chisel developers heavily rely on bundles to keep the design organized and readable. However, previously only the entire bundle could be marked as faulty due to the way the transforms were implemented. \textsc{Nail} allows users to target specific (sub)signals within a bundle for fault injection. This can significantly reduce the scan chain size and make fault configuration more readable. For example, consider a bundle with one control bit and $n$ data bits. With \textsc{Nail}, the control bit can be marked as faulty while leaving the data bits intact, saving $n$ bits in the scan chain configuration. 
\section{Experimental Results} \label{sec:exper}
To demonstrate the enhanced controllability and usability enabled by \textsc{Nail}, we present a fault injection experiment that cannot be realized with state-of-the-art tools without extensive manual modification to the Chisel design.
 
The experiment targets the general-purpose register file in a 64-bit Rocket Chip~\cite{RocketChip}, a Chisel-based RISC-V processor. The register file is implemented as a \texttt{chisel3.Mem} and represented in Chirrtl~\cite{chirrtl} (a dialect of FIRRTL) as \verb|cmem rf : UInt<64> [31]|. At this stage of the design flow, the register file is still treated as a black box, with internal implementation details not yet determined.
Therefore, it is not possible to mark only some parts inside the memory as faulty, as the memory technology is not yet specified. The approach demonstrated here can therefore model faults in \textit{any memory} structure and, by extend, any black box.

The goal is to model a fault in a single, runtime-selectable general-purpose register. Fig.~\ref{fig:faultyRegfileExp} shows a schematic representation of this experiment, highlighting the functional blocks within the Rocket chip, the scan chain integration and the RoCC controller. Fig.~\ref{fig:faultyRegfileScala} presents a Chisel code snippet from the Rocket module, illustrating the modification needed to enable fault injection on the register file's write data (\texttt{rf\_wdata}). Here, the injection condition is based on the targeted register address (\texttt{rf\_waddr}) and the write enable signal (\texttt{rf\_wen}).

\begin{figure}[htbp]
    \tikzmath{
    \conditionerPadX = 1;
    \conditionerPadY = .5;
    \componentDx = 1.5;
    \componentDy = 1.5;
}

\begin{tikzpicture}[
    pics/ControllerRFP/.style={
        code = {
            \draw [RegS, rounded corners = .25cm] (0, 0) rectangle ++(\componentDx, -1.5*\componentDy);
        }
    },
    pics/ConditionerRFP/.style={
        code = {
            \coordinate (condIoIn) at (0, -.5*\componentDy);
            \coordinate (condIoOut) at (\componentDx, -.5*\componentDy);
            \draw [RegS, rounded corners = .25cm] (0, 0) rectangle ++(\componentDx, -\componentDy);
        }
    }, 
    pics/InjectorRFP/.style={
        code = {
            \coordinate (injIoIn) at (0, -.5*\componentDy);
            \coordinate (injIoOut) at (\componentDx, -.5*\componentDy);
            \draw [RegS, rounded corners = .25cm] (0, 0) rectangle ++(\componentDx, -\componentDy);
        }
    },
    pics/RegFileRFP/.style={
        code = {
            \draw [RegS, rounded corners = .25cm] (0, 0) rectangle ++(\componentDx, -1.5*\componentDy);
        }
    }, 
    pics/RocketRFP/.style={
        code = {
            \draw [RegS, rounded corners = .25cm] (0, 0) rectangle ++(4*\componentDx, -3*\componentDy);
        }
    }, 
    pics/AddrRFP/.style={
        code = {
            \draw [RegS, rounded corners = .25cm] (0, 0) rectangle ++(1.1*\componentDx, -.5*\componentDy);
        }
    }, 
    pics/DataRFP/.style={
        code = {
            \draw [RegS, rounded corners = .25cm] (0, 0) rectangle ++(1.1*\componentDx, -.5*\componentDy);
        }
    }
]
\pic [local bounding box=rocket] at (5, 0) {RocketRFP};
\pic [local bounding box=fc] at ($(rocket.east) - (7.75, 0)$) {ControllerRFP};
\pic [local bounding box=cond] at ($(fc.north east) + (.5, -.5)$) {ConditionerRFP};
\pic [local bounding box=inj] at ($(cond.north east) + (.5, 0)$) {InjectorRFP};
\pic [local bounding box=rf] at ($(inj.north east) + (.5, .5*\componentDy)$) {RegFileRFP};
\pic [local bounding box=data] at ($(rocket.north west) + (2, -1.5)$) {DataRFP};
\pic [local bounding box=addr] at ($(rocket.north west) + (2, -.5)$) {AddrRFP};

\draw[InBetweenConnect2S] (injIoOut) -- ($(rf.west |- injIoOut)$);
\draw[InBetweenConnect2S] ($(data.south -| inj.north)$) -- (inj.north);
\draw[InBetweenConnect1S] (addr.180) -| (cond.north);
\draw[InBetweenConnect1S] (addr) -| (rf.north);

\draw[ScanChainS] (condIoOut) -- ($(cond.west -| injIoIn)$);
\draw[ScanChainS] ($(fc.east |- condIoIn.west)$) -- (condIoIn);

\node[font=\small, anchor=center, rotate=-45] at ($(cond.center)$) {Conditioner};
\node[font=\small, anchor=center, rotate=-45] at ($(inj.center)$) {Injector};
\node[font=\small, align=center, anchor=center] at ($(fc.center)$) {Software-\\controlled\\Coprocessor};
\node[font=\small, align=center, anchor=center] at ($(rf.center)$) {Register\\File};

\node[font=\small, align=center, anchor=center] at ($(data.center)$) {\texttt{Data}};
\node[font=\small, align=center, anchor=center] at ($(addr.center)$) {\texttt{Address}};

\node[anchor=east] at ($(rocket.north west) + (1.5, -.5)$) {Rocket};

\end{tikzpicture}
    \caption{
    Schematic of the fault injection experiment driven by a software controlled RoCC core.
    Red: The (faulty) data to be stored in the register file (in code \texttt{rf\_wdata}).  
    Blue: The write address (in code \texttt{rf\_waddr}) compared to the target value.
    Green: The scan chain driven by the controller.
    }\label{fig:faultyRegfileExp}
\end{figure}

\subsection{Experiment Workflow}
The procedure is divided into two main parts: 
\emph{(1) Configuring the scan fields and loading the configuration} and
\emph{(2) Enabling the scan chain and verifying the correct injection of the fault (the driver code)}.

\begin{itemize}
    \item \textbf{Configuring the Fault:} 
    Using the header emitted by the \textsc{NailCompanion}, we have access to the scan chain present in the design. First, we specify the target address by setting the corresponding scan field, then activate the conditioner and injector by setting their respective \texttt{isActive} fields.
    Next, we configure the mask to indicate which bits should be modified (here, all bits of the data being written to memory), and specify the value to inject into the masked bits. Finally, we prepare the scan chain and send the underlying array to the coprocessor.
    \item \textbf{Checking the Fault:} 
    The driver code validates whether the fault occurs.
    First, the scan chain is enabled by sending an instruction to the coprocessor.
    We then write to the targeted register and read back from it. If the read value matches the injected fault, the fault was successfully injected, and the scan chain is subsequently disabled.
\end{itemize}

An example implementation of this process in C++ using the \textsc{NailCompanion} is shown in Fig.~\ref{fig:FaultyRegfileCpp}. The first part of configuring the scan chain is done in the \texttt{load\_fault} function, the second part is done by \texttt{check\_target}.

\begin{figure}[htbp]
    \centering
    \lstset{language=Scala,
    basicstyle=\footnotesize\ttfamily , keywordstyle=\color{UzLBlue_medium}\footnotesize , stringstyle=\color{UzLRed}\footnotesize ,
    commentstyle=\color{UzLGreen}\footnotesize ,
    morecomment=[l][\color{UzLRed}]{\#}
    }
    \begin{lstlisting}
class Rocket(tile: RocketTile)
(implicit p: Parameters) extends CoreModule()(p)
    with HasRocketCoreParameters
    with HasRocketCoreIO
    with nail.ContainsCondition
    with nail.NailInjectee {
    // ...
    // Register file write address (5 bits)
    val rf_waddr = ...
    // Register file write data (XLEN bits)
    val rf_wdata = ...
    
    // MODIFICATIONS BY NAIL:
    private val targetAddr = CondScanField(5)
    
    // Condition: match targetAddr and write en.
    private val faultyWrite = 
    (targetAddr === rf_waddr) && rf_wen

    // Fault rf_wdata only if faultyWrite is 1 
    markAsFaulty(rf_wdata,
                "rocket", 
                Some(classOf[StuckAtInjectorN]),
                faultyWrite)
    // ...    
}
    \end{lstlisting}
    \caption{
        This Scala code snippet from the Rocket core illustrates fault injection applied to the register file write data (\texttt{rf\_wdata}), achieved by adding only three lines of code to the design. The injection is triggered by comparing the register file write address (\texttt{rf\_waddr}) with a runtime-configurable target address. This facilitates modeling faults such as a faulty lane in memory.
    }\label{fig:faultyRegfileScala}
\end{figure}

\begin{figure}
    \centering
    \lstset{language=C++,
    basicstyle=\footnotesize\ttfamily , keywordstyle=\color{UzLBlue_medium}\footnotesize , stringstyle=\color{UzLRed}\footnotesize ,
    commentstyle=\color{UzLGreen}\footnotesize ,
    morecomment=[l][\color{UzLRed}]{\#}
    }
    \begin{lstlisting}
#include "nail-companion.hpp"
using namespace nail;
using namespace nail::types;
using reg_t = uint64_t; // XLEN is 64 bits

// Mask out all bits in reg_t
constexpr reg_t mask =
    std::numeric_limits<reg_t>::max();
// The value to inject into the register
constexpr reg_t TARGET = 0xC0FFEE;
// Contains the "rocket" scan chain repr.
static sc::rocket::scan_chain_t chain{};

// The driver code:
void check_target() {
    // Init x15 to 0
    volatile int var __asm__("x15") = 0;
    // Enable the fault
    send_RoCC<sc::rocket::nail_gun, f_ENABLE>();
    // Wait for the fault to take effect
    asm volatile (
        "1:\n"
        "li t0, %1\n"
        "mv t1, %0\n"
        "bne t1, t0, 1b\n"
        :
        : "r"(var), "i"(TARGET)
        : "t0", "t1"
    );
    // Disable the fault
    send_RoCC<sc::rocket::nail_gun, f_ENABLE>();
}
// Config. the scan chain. Set x15 to TARGET
void load_fault(){
    // Fault x15 register
    chain.set<Rocket::rf_wdata,
        conditions::Condition>(15);
    chain.set<Rocket::rf_wdata,
        conditions::IsActive>(true);
    // Enable injector
    chain.set<Rocket::rf_wdata,
        inject::IsActive>(true);
    // Mask out all bits
    chain.set<Rocket::rf_wdata,
        inject::StuckAtMask>(mask);
    // Set them to TARGET
    chain.set<Rocket::rf_wdata,
        inject::StuckAt>(TARGET);
    // Gen. checksum and send configuration
    chain.prepare();
    send_RoCC<sc::rocket::nail_gun, f_LOAD>(&chain);
}
int main() {
    load_fault();
    check_target();
}
    \end{lstlisting}
    \caption{C++ code using \textsc{NailCompanion} to configure and verify register file fault injection in the Rocket core. The scan chain is set to inject a specific value (\texttt{TARGET}) into register \texttt{x15}.}\label{fig:FaultyRegfileCpp}
\end{figure}

\subsection{Simulation and Emulation Results}
\textsc{Nail}'s approach was validated both in simulation (using Verilator) and on an FPGA. Table~\ref{tab:resourceUtilization}, summarizes the resource utilization for the baseline and instrumented designs. The setup for design synthesis is as follows:
\begin{itemize}
    \item Target device: Xilinx ZCU104 FPGA
    \item Synthesis tool: Vivado 2024.2
    \item Rocket Chip commit: f517abb
    \item Chisel: \texttt{chisel3} at 3.6.1 
\end{itemize}

\begin{table}[htbp]
    \centering
    \begin{tabularx}{\columnwidth}{lXX}
        \hline
        \textbf{Resource} & \textbf{Baseline} & \textbf{With} \textsc{Nail}  \\  \hline
        LUT (\%) & 26.43 & 26.49 \\ \hline
        LUTRAM (\%) & 4.39 & 4.61 \\ \hline
        FF (\%) & 9.59 & 9.67 \\ \hline
    \end{tabularx}
    \caption{Resource utilization for the Rocket core with and without fault injection. Both were synthesized with 100Mhz.
    }\label{tab:resourceUtilization}
\end{table}

As the results indicate, the proposed approach incurs less than 1\% additional hardware resource usage and introduces no timing penalty, while preserving compatibility with standard Chisel3 toolchains and RISC-V compilers. This demonstrates that \textsc{Nail} enables advanced, fine-grained, and runtime-configurable fault injection into memory black boxes with minimal impact on design resources and performance.
\section{Conclusion}\label{sec:con}
This work introduces \textsc{Nail}, an open-source FI framework for Chisel-generated hardware designs.
Existing simulation-based FI tools offer fine-grained control but are often slow and can be difficult to use. Emulation-based tools provide high speed but lack controllability which makes many fault scenarios impossible to analyze.

\textsc{Nail} overcomes this limitation by introducing state-based triggers that can be configured at runtime. Combined with state-of-the-art techniques, this enables a wide range of fault trigger controls. These range from coarse-grained instruction-level triggers, when low granularity suffices, to highly granular triggers that compare internal states against runtime-controlled parameters.
To interact with the instrumented design and to generate new fault parameters at runtime, \textsc{Nail} provides its own C++-based interface, the \textsc{NailCompanion}. It uses modern C++20 features to enable readable and efficient control over the fault injection experiment.

We demonstrate \textsc{Nail}'s features by modeling a fault in a general-purpose register within a RISC-V core—a scenario previously impossible to model—and validate our approach both in simulation and FPGA emulation. Our results show that this can be achieved with less than 1\% resource overhead and as little as three additional lines of code in the Chisel design.

Collectively, these advancements position \textsc{Nail} as a robust and versatile tool for fault injection in digital hardware designs.

\section*{Acknowledgment}

\bibliography{bib}

\end{document}